\documentclass[,,preprintnumbers,amsmath,amssymb]{revtex4}
\usepackage{graphicx}
\usepackage{dcolumn}
\usepackage{bm}
\usepackage{hyperref}
\usepackage{array}

\begin{document}

\title{Coupling Constant $g_{\Lambda_{b}B_{s}\Lambda}$ using the Baryonic $\bar{B}^{0}_{s}\rightarrow p\bar{\Lambda} K^{-}$ Decay}

\author{S. Rafibakhsh}%
\author{H. Mehraban}

\email[]{Rafibakhsh@semnan.ac.ir (corresponding author)}
\email[]{Hmehraban@semnan.ac.ir }%
\affiliation {Physics Department, Semnan University, P.O.Box 35195-363, Semnan, Iran}

\begin{abstract}
We study the three-body baryonic B decay $\bar{B}^{0}_{s}\rightarrow p\bar{\Lambda} K^{-}$ within the framework of the pole model via the baryonic $\Lambda_{b}$ pole. In our calculation, we require the strong coupling constant $g_{\Lambda_{b}B_{s}\Lambda}$ and investigate if $g_{\Lambda_{b}B_{s}\Lambda}=10.49\pm1.57$ is adopted, the branching ratio agrees with the experimental result, reported by the LHCb collaboration.
\end{abstract}

\maketitle
\section{Introduction}
\label{{sec:1}}

Since the late 1980’s when the first three- and four-body baryonic B decay modes $p\bar{p}\pi^{\pm}$ and $p\bar{p}\pi^{+}\pi^{-}$ were observed by ARGUS ~\cite{1}, rich studies in this field started by calculating the branching ratios, angular and CP asymmetries ~\cite{2}. Even though the first claim of ARGUS was soon refuted by CLEO collaboration ~\cite{3}, it stimulated new baryonic observations and led to thriving theoretical baryonic B decay studies in the last decades, which are still an active field in phenomenology. As exact QCD calculations are so difficult, different theoretical models were employed to calculate B decay rates e.g. pole model ~\cite{4,5}, diquark model ~\cite{6}, and QCD sum rules ~\cite{7,8,9}.

In the pole model, first applied by Deshpande ~\cite{4} and Jarfi ~\cite{5} and then developed by Cheng and Yang ~\cite{10,11}, an intermediate b-flavored hadron state is considered which then decays to the final state hadrons. This model has been successful in calculating several baryonic B decay rates, e.g. $B^{-}\rightarrow\Sigma^{0}_{c}\bar{p}$, $\bar{B}^{0}\rightarrow \Lambda^{+}_{c}\bar{p}$, $B^{0}\rightarrow\Sigma^{++}_{c}\bar{p}\pi^{-}$ ~\cite{10} and $\bar{B}^{0}\rightarrow p\bar{n}\pi^{-}$~\cite{12}.

The present paper applies the pole model to study the first decay of $B_{s}^{0}$ observed, which is $\bar{B}^{0}_{s}\rightarrow p\bar{\Lambda} k^{-}$. This decay has been studied in ~\cite{13} by Geng and Hou, and their rate obtained is in excellent agreement with the LHCb collaboration measurement ~\cite{14}. Although their approach is the most accurate one in calculating the branching ratios so far, the pole model can lead us to guess the unknown coupling constants.

\section{Formalism}
\label{sec:2}
\begin{figure}
\includegraphics[trim=0 16.5cm 1cm 1.5cm, scale=0.8, clip]{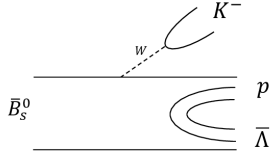}
\caption{Quark diagram for $\bar{B}^{0}_{s}\rightarrow p\bar{\Lambda} K^{-}$},
\label{fig:i}
\end{figure}

Based upon the quark diagrams of the decay $\bar{B}^{0}_{s}\rightarrow p\bar{\Lambda} K^{-}$ depicted in Fig. ~\ref{fig:i}, the tree-level dominated amplitude under the factorization hypothesis is written as
\begin{equation}
\label{eq:1}
\mathcal{M }= \frac{G_F}{\sqrt{2}} V_{ub} V^*_{us}\, a_1 \langle \bar{\Lambda}p|(\bar{u}b)_{V-A}|\bar{B}^{0}_{s}\rangle
\langle K^{-}|(\bar{s} u)_{V-A}|0\rangle,
\end{equation}
where $G_F$ is the Fermi constant, $V_{ij}$ is the CKM matrix element, $(\bar{q}_1 q_2)_{V-A}=\bar{q}_1\gamma_\mu (1-\gamma_5 )q_2$ and the parameter $a_1=c_1^{eff}+c_2^{eff}/N_c^{eff}$ with the effective Wilson coefficients $c_i^{eff}$ and the effective color number $N_c^{eff}$ ranging between 2 and $\infty$, which will be discussed later.
The mesonic matrix element is written as ~\cite{15}
\begin{equation}
\label{eq:2}
\langle K^{-}|(\bar{s} u)_{V-A}|0\rangle=if_{K}p_{\mu}
\end{equation}
with $f_K$ the decay constant.

\begin{figure}
 \includegraphics[trim=0 15.5cm 1cm 1.5cm, scale=0.8, clip]{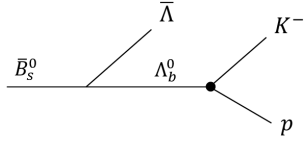}
\caption{Pole diagram for $\bar{B}^{0}_{s}\rightarrow p\bar{\Lambda} K^{-}$ where the black dot represents the weak vertex}
\label{fig:j}
\end{figure}

To evaluate the three-body baryonic matrix element $\langle \bar{\Lambda}p|(\bar{u}b)_{V-A}|\bar{B}^{0}_{s}\rangle$ using the pole model, the intermediate state $\Lambda_{b}$ is considered which leads to the strong decay $\bar{B}_{s}^{0}\rightarrow\bar{\Lambda}\Lambda_{b}$ followed by the weak decay $\Lambda_{b}\rightarrow K^{-}p$ as depicted in Fig.~\ref{fig:j}. To apply factorization to the weak decay $\Lambda_{b}\rightarrow K^{-}p$, the heavy-light baryon form factor is needed as follows~\cite{12}
\begin{eqnarray}
\label{eq:3}
&&\langle p(p_{p})|(\bar{u}b)_{V-A}|\Lambda_{b}(p_{\Lambda_{b}})\rangle\cr \cr
&=&\bar{u}_{p}\{f_{1}^{\Lambda_{b}p}(p_{K}^{2})\gamma_{\mu}+i\frac{f_{2}^{\Lambda_{b}p}(p_{K}^{2})}{m_{\Lambda_{b}}+m_{p}}\sigma_{\mu\nu}p_{K}^{\nu}
+\frac{f_{3}^{\Lambda_{b}p}(p_{K}^{2})}{m_{\Lambda_{b}}+m_{p}}p_{K\mu}\cr
&\pm&[g_{1}^{\Lambda_{b}p}(p_{K}^{2})\gamma_{\mu}+i\frac{g_{2}^{\Lambda_{b}p}(p_{K}^{2})}{m_{\Lambda_{b}}+m_{p}}\sigma_{\mu\nu}p_{K}^{\nu}
+\frac{g_{3}^{\Lambda_{b}p}(p_{K}^{2})}{m_{\Lambda_{b}}+m_{p}}p_{K\mu}]\gamma_{5}
\}u_{\Lambda_{b}},\qquad
\end{eqnarray}
where $p_{K}=p_{\Lambda_{b}}-p_{p}$
and $(f,g)_{i}^{\Lambda_{b}p}$ are the heavy-light form factors that are evaluated by the non-relativistic quark model ~\cite{12,16,17} and the decay amplitude is obtained as follows
\begin{equation}
\label{eq:4}
\mathcal{M }=\frac{G_F}{\sqrt{2}}\,\alpha\,\beta\,\bar{u}_{p}\{A{\not}p_{K}
+B{\not}p_{K}\gamma_{5}
+C\gamma_{5}+D
\}v_{\bar{\Lambda}},
\end{equation}
where $\alpha=1/(p_p+p_K)^{2}-m^{2}_{\Lambda_{b}}$, $\beta=V_{ub} V^*_{us}\, a_1\,f_K\,g_{\Lambda_{b}B_{s}\Lambda}$ with $g_{\Lambda_{b}B_{s}\Lambda}$ being the coupling, and
\begin{eqnarray}
\label{eq:5}
A&=&f_{1}\,g_{1}(m_{\Lambda_{b}}+m_{p}),\qquad\qquad\,
B=-f_{1}(m_{\Lambda_{b}}-m_{p}),\cr
C&=&-f_{1}(s-m_{p}^{2}-m_{K}^{2}),\qquad\quad
D=-f_{1}g_{1}(s-m_{p}^{2}-m_{K}^{2}),
\end{eqnarray}
where the Dalitz plot analysis is applied, in which the invariants are defined by~\cite{18}
\begin{eqnarray}
\label{eq:6}
s=(p_{\Lambda}+p_{K})^{2}=(p_{B_s}-p_{p})^{2},\qquad t=(p_{\Lambda}+p_{p})^{2}=(p_{B_s}-p_{K})^{2}\,.
\end{eqnarray}
Now using the Casimir trick, the amplitude’s absolute square is obtained as follows
\begin{eqnarray}
\label{eq:7}
\sum|M|^{2}&=&G_{F}^{2}\,\{|A|^{2}\,[(m_{B_s}^{2}+m_{\Lambda}^{2}-t-s)\,(s-m_{\Lambda}^{2}-m_{K}^{2})\cr\cr
&-&m_{K}^{2}\,(t-(m_{\Lambda}-m_{p})^{2})] \cr \cr
&+&2\,\,|AD|[m_{p}(s-m_{\Lambda}^{2}-m_{K}^{2})\cr\cr
&-&m_{\Lambda}\,(m_{B_s}^{2}+m_{\Lambda}^{2}-t-s)]\}\cr\cr
&+&|D|^{2}[t-(m_{\Lambda_c}-m_p)^{2}]\cr\cr
&+&(A\rightarrow B),(D\rightarrow C),(m_{\Lambda}\rightarrow -m_{\Lambda}).
\end{eqnarray}
The decay width for a three-body decay is computed by
\begin{eqnarray}
\label{eq:8}
\Gamma=\frac{1}{(2\pi)^{3}}\,\frac{1}{32\,m_{B_s}^{3}}\int\int(\,\sum|M|^{2}\,)\,ds\,dt\,,
\end{eqnarray}
where the boundaries are taken as follows
\begin{eqnarray}
\label{eq:9}
s_{min}=(m_{p}+m_{\Lambda})^{2},\qquad s_{max}=(m_{B_s }-m_{K})^{2}\,,
\end{eqnarray}
and
\begin{eqnarray}
\label{eq:10}
t_{\stackrel{\hbox{\emph{max}}}{\hbox{\emph{min}}}}&=&m_{\Lambda}^{2}+m_{p}^{2}-\frac{1}{s}\,[\,(s-m_{B_s}^{2}+m_{K}^{2})\,(s+m_{\Lambda}^{2}-m_{p}^{2}) \cr \cr
&\pm&\lambda^{1/2}(\,s\,,\,m_{B_s}^{2}\,,\,m_{K}^{2}\,)\,\lambda^{1/2}(\,s\,,\,m_{\Lambda}^{2}\,,\,m_{p}^{2}\,)\,]\,,
\end{eqnarray}
with
\begin{eqnarray}
\label{eq:11}
\lambda(a,b,c)=a^{2}+b^{2}+c^{2}-2\,(ab+ac+bc)\,.
\end{eqnarray}
Ultimately, the branching ratio is calculated as the follwing
\begin{eqnarray}
\label{eq:12}
BR=\frac{\Gamma(\bar{B}^{0}_{s}\rightarrow p\bar{\Lambda} K^{-})}{\Gamma_{tot}}\,
\end{eqnarray}

\section{Numerical Results}
\label{sec:2}

For the numerical analysis, we adopt the CKM matrix elements, the hadron masses ~\cite{14}, the decay constant $f_k$ ~\cite{13}, and the form factors $(f,g)_{1}^{\Lambda_bp}(p_k^{2})$ ~\cite{12,17} as follows
\begin{eqnarray}
\label{eq:13}
&V_{ub}=(3.48\pm 0.4\pm 0.07 )\times10^{-3},\qquad V_{us}=0.2245\pm 0.0008, \cr\cr
&m_{B_{s}^{0}}=5366.92\pm 0.10 Mev,\qquad m_{\Lambda}=1115.683\pm0.006 Mev, \cr\cr
&m_p=938.272\pm0.00000029 Mev,\qquad m_{k^{-}}=493.677\pm0.016 Mev, \cr\cr
&m_{\Lambda_b}=5619.60\pm0.17 Mev,\cr\cr
&f_k=156.2\pm0.7 Mev,\cr\cr
&f_{1}^{\Lambda_bp}(p_k^{2})=f_{1}^{\Lambda_bp}(p_k^{2})=0.86.
\end{eqnarray}

In addition, the parameter $a_1$ may be strongly affected by the nonfactorizable effects. Therefore, it is customarily considered a free parameter, and its value is obtained from the data, using different form-factor models. Various values for $a_1$ within different models are listed in ~\cite{19,20}. We consider $a_1=0.8$, as an illustration, to achieve the best result in our work.
Putting everything together, the branching ratio is numerically obtained
\begin{eqnarray}
\label{eq:14}
BR(\bar{B}^{0}_{s}\rightarrow p\bar{\Lambda} K^{-})=5.0\times10^{-8}g_{\Lambda_{b}B_{s}\Lambda}^{2}
\end{eqnarray}

Comparing this result with the experimental measurement reported by the LHCB collaboration $BR_{Exp}(\bar{B}^{0}_{s}\rightarrow p\bar{\Lambda} K^{-})=(5.5\pm1.0)\times10^{-6}$, we conclude if the coupling takes the value of $g_{\Lambda_{b}B_{s}\Lambda}=10.49\pm1.57$, the branching ratio is consistent with the experimental result. This coupling is not available in the literature. Nonetheless, the value of $g_{\Lambda_{b}BN}=12.67\pm3.76$ is obtained in Ref.~\cite{21} using the QCD sum rules, by Azizi et al. Taking into account the SU(3) flavor symmetry breaking effects, the values for strong coupling constants $g_{\Lambda_{b}B_{s}\Lambda}$ and $g_{\Lambda_{b}BN}$ are expected to be in the same order. Hence, the obtained result in the present study is reasonable.

\section{Conclusion}
\label{sec:3}
We have studied the three-body baryonic B decay $\bar{B}^{0}_{s}\rightarrow p\bar{\Lambda} K^{-}$ utilizing the pole model approach. We take the baryon $\Lambda_b$ as the intermediate state that leads to a strong and a weak decays: $\bar{B}_{s}^{0}\rightarrow\bar{\Lambda}\Lambda_{b}$ and $\Lambda_{b}\rightarrow K^{-}p$. In the process of calculating the branching ratio, we need the strong coupling constant $g_{\Lambda_{b}B_{s}\Lambda}$ and find if we employ $g_{\Lambda_{b}B_{s}\Lambda}=10.49\pm1.57$, the branching ratio is equal to $(5.5\pm1.0)\times10^{-6}$, which is the result of LHCb collaboration. To validate our obtained value for $g_{\Lambda_{b}B_{s}\Lambda}$, we compare it with the existing coupling $g_{\Lambda_{b}BN}$. Since the coupling of these two vertices is numerically of the same order and close to each other considering the SU(3) flavor symmetry breaking effects, we consider our obtained value to be a reasonabe result.

\section*{Acknowledgements}
The authors would like to express their gratitude to Prof. K. Azizi for his helpful discussion on comparison of coupling constants.

\end{document}